\documentclass[a4paper,12pt]{article}
\usepackage{amsmath}
\usepackage[psamsfonts]{amssymb}
\usepackage{rsfs}
\usepackage{bm}

\usepackage{cite}
\usepackage[dvips]{graphicx}
\usepackage{color}

\makeatletter
\@addtoreset{equation}{section}
\makeatother

\begin{document} 

\begin{titlepage}

\baselineskip 10pt
\hrule 
\vskip 5pt
\leftline{}
\leftline{Chiba Univ./KEK Preprint
          \hfill   \small \hbox{\bf CHIBA-EP-168}}
\leftline{\hfill   \small \hbox{\bf KEK Preprint 2008-1}}
\leftline{\hfill   \small \hbox{September 2008}}
\vskip 5pt
\baselineskip 14pt
\hrule 
\vskip 1.0cm
\centerline{\Large\bf 
New descriptions of 
} 
\vskip 0.3cm
\centerline{\Large\bf  
lattice SU(N) Yang-Mills theory
}
\vskip 0.3cm
\centerline{\Large\bf  
towards quark confinement 
}
\vskip 0.3cm
\centerline{\large\bf  
}

\vskip 0.5cm

\centerline{{\bf 
Kei-Ichi Kondo$^{\dagger,\ddagger,{1}}$,  
Akihiro Shibata$^{\flat,{2}}$, 
Toru Shinohara$^{\ddagger,{3}}$,
}}
\vskip 0.3cm
\centerline{{\bf 
Takeharu Murakami$^{\ddagger,{4}}$,
Seikou Kato$^{\sharp,{5}}$, 
$\&$
Shoichi Ito$^{\star,{6}}$
}}  
\vskip 0.5cm
\centerline{\it
${}^{\dagger}$Department of Physics, Graduate School of Science, 
Chiba University, Chiba 263-8522, Japan
}
\vskip 0.3cm
\centerline{\it
${}^{\ddagger}$Graduate School of Science and Technology, 
Chiba University, Chiba 263-8522, Japan
}
\vskip 0.3cm
\centerline{\it
${}^{\flat}$Computing Research Center, High Energy Accelerator Research Organization (KEK)   
}
\vskip 0.1cm
\centerline{\it
\& 
Graduate Univ. for Advanced Studies (Sokendai), 
Tsukuba 
305-0801, 
Japan
}
\vskip 0.3cm
\centerline{\it
${}^{\sharp}$Takamatsu National College of Technology, Takamatsu 761-8058, Japan
}
\vskip 0.3cm
\centerline{\it
${}^{\star}$Nagano National College of Technology, 716 Tokuma, Nagano 381-8550, Japan
}

\centerline{\it
}
\vskip 0.5cm

\begin{abstract}
We give new  descriptions of lattice SU(N) Yang-Mills theory in terms of new lattice variables. 
The validity of such descriptions has already been demonstrated in the $SU(2)$ Yang-Mills theory by our previous works  from the viewpoint of defining and extracting  topological degrees of freedom such as gauge-invariant magnetic monopoles and vortices which play the dominant role in quark confinement.  
In particular,  we have found that the SU(3) lattice Yang-Mills theory has two possible options,  maximal and minimal:  
The existence of the minimal option has been overlooked so far, while the maximal option reproduces the conventional SU(3) Cho-Faddeev-Niemi-Shabanov decomposition in the naive continuum limit.
The new description gives an important framework for  understanding the mechanism of quark confinement based on the dual superconductivity.

\end{abstract}


\hrule  
\vskip 0.1cm
  E-mail:  
${}^1$ 
{\tt kondok@faculty.chiba-u.jp};
${}^2$ 
{\tt akihiro.shibata@kek.jp};

${}^3$ 
{\tt sinohara@graduate.chiba-u.jp};
${}^4$ 
{\tt tom@fullmoon.sakura.ne.jp}
  
${}^5$ 
{\tt kato@takamatsu-nct.ac.jp};
${}^6$ 
{\tt shoichi@ei.nagano-nct.ac.jp};

\par 
\par\noindent


\vskip 0.5cm

\newpage




\end{titlepage}


\pagenumbering{arabic}

\baselineskip 14pt
\section{Introduction}


The purpose of this Letter is to give new descriptions  of the Yang-Mills theory \cite{YM54} on a lattice, which are expected to give an efficient framework to explain  quark confinement based on the dual superconductivity picture.  
The dual superconductivity \cite{dualsuper} is conjectured to occur due to the condensation of magnetic monopoles, just as the ordinary superconductivity is caused by the condensation of the Cooper pairs.  
In the dual superconductor, the dual Meissner effect forces the color electric flux between a quark and an antiquark to be squeezed into a tube like region forming the hadronic string.

In view of these, we wish to construct new descriptions which enable us to extract the dominant degrees of freedom that are relevant to quark confinement according to the Wilson criterion in such a way that they reproduce almost all the string tension of the linear inter-quark potential.
In a previous paper \cite{KSM08}, we have considered the decomposition in which the original $SU(N)$ Yang-Mills field $\mathscr{A}_\mu(x)$ is decomposed into two parts, $\mathscr{V}_\mu(x)$ and $\mathscr{X}_\mu(x)$, i.e., 
\begin{equation}
 \mathscr{A}_\mu(x)=\mathscr{V}_\mu(x)+\mathscr{X}_\mu(x)
  ,
  \label{decomp1}
\end{equation}
such that  
$\mathscr{V}_\mu$ transforms under the gauge transformation just like the original gauge field $\mathscr{A}_\mu$, while $\mathscr{X}_\mu$ transforms like an adjoint matter field:
\begin{subequations}
\begin{align}
  \mathscr{V}_\mu(x) & \rightarrow \mathscr{V}_\mu^\prime(x) = \Omega(x) (\mathscr{V}_\mu(x) + ig^{-1} \partial_\mu) \Omega^{-1}(x) 
 ,
 \label{V-ctransf}
  \\
  \mathscr{X}_\mu(x) & \rightarrow \mathscr{X}_\mu^\prime(x) = \Omega(x)  \mathscr{X}_\mu(x) \Omega^{-1}(x) 
 \label{X-ctransf}
 . 
\end{align}
In the decomposition (\ref{decomp1}), a crucial role is played by a unit vector field $\bm{n}(x)$ called the \textit{color field}.
The color field is defined by the following property.
It must be a functional or composite operator of the original Yang-Mills field  $\mathscr{A}_\mu(x)$ such that it transforms under the gauge transformation according to the adjoint representation:  
\begin{align}
  \bm{n}(x) & \rightarrow \bm{n}^\prime(x) = \Omega(x)  \bm{n}(x) \Omega^{-1}(x) 
   .
 \label{n-ctransf}
\end{align}
\end{subequations}
The prescription for obtaining such a color field is to be specified separately.  
The usefulness of the color field is as follows.  
It enables one to give a gauge-invariant definitions of magnetic monopole with the integer-valued magnetic charges subject to a quantization condition analogous to the Dirac type. 
For $SU(2)$, indeed, once such a \textit{single} color field $\bm{n}(x)$ of unit length is introduced, we can define
the \textit{gauge-invariant} two-form by 
\begin{align}
 f_{\alpha\beta}(x) 
  :=  2{\rm tr}\{ \partial_\alpha [\bm{n}(x) \mathscr{A}_\beta(x)]  - 
  \partial_\beta  [\bm{n}(x)  \mathscr{A}_\alpha(x) ] 
+  ig^{-1}  \bm{n}(x) [  \partial_\alpha \bm{n}(x) ,  \partial_\beta \bm{n}(x)]  \}
   ,
\end{align}
and the gauge-invariant ``magnetic-monopole current''   $k$  by
\begin{align}
 k_\mu = \frac12 \epsilon_{\mu\nu\alpha\beta} \partial_\nu f_{\alpha\beta}
 ,
\end{align}
which is conserved in the sense that 
$\partial_\mu k_\mu=0$.
Consequently, the magnetic charge $q_m$ is gauge-invariant and obeys the quantization condition: 
\begin{align}
 q_m :=\int d^3 \tilde{\sigma}_\mu k_\mu 
 = 4\pi g^{-1} n, \quad n \in \mathbb{Z} := \{ 0, \pm 1, \pm 2, \cdots  \} 
  ,
\end{align}
where $\bar{x}^\mu$ denotes a parameterization of the 3-dimensional volume $V$ and 
 $d^3 \tilde{\sigma}_{\mu} $ is the dual of the 3-dimensional volume element $d^3 \sigma^{\gamma_1\gamma_2\gamma_3}$.

 For $SU(2)$,  $\mathscr{V}_\mu$ and $\mathscr{X}_\mu$ are specified by two defining equations by way of the color field:
\begin{subequations}
\begin{align}
 0 =&  \mathscr{D}_\mu[\mathscr{V}] \bm{n}(x) :=  \partial_\mu \bm{n}(x)-ig [ \mathscr{V}_\mu(x), \bm{n}(x) ]   ,
\\
  0 =& {\rm tr}[\mathscr{X}_\mu(x)  \bm{n}(x) ] 
 . 
\end{align}
\end{subequations}
In fact, by solving these defining equations, $\mathscr{V}_\mu(x)$ and $\mathscr{X}_\mu(x)$ in $SU(2)$ are determined uniquely as functions of $\bm{n}(x)$ and $\mathscr{A}_\mu(x)$:
\begin{subequations}
\begin{align}
  \mathscr{V}_\mu(x) =&   2{\rm tr}[\mathscr{A}_\mu(x) {\bm n}(x)] {\bm n}(x)   +i g^{-1}  [ {\bm n}(x) , \partial_\mu {\bm n}(x) ]
 ,
\\
  \mathscr{X}_\mu(x) =&   -ig^{-1}  [\bm{n}(x) , \mathscr{D}_\mu[\mathscr{A}]\bm{n}(x) ]
 .  
\end{align}
 \label{V-X}
\end{subequations}

 From these point of view, we have already given a new framework for $SU(2)$ Yang-Mills theory \textit{on a lattice} and subsequently presented numerical evidences supporting the validity of the new lattice framework by performing numerical simulations \cite{KKMSSI05,IKKMSS06,SKKMSI07}.
The relevant lattice description could be regarded as a lattice version of the continuum formulation \cite{KMS06} which was known as the Cho-Faddeev-Niemi-Shabanov (CFNS) decomposition \cite{Cho80,FN98,Shabanov99}. 
In previous papers \cite{KKMSSI05,IKKMSS06,SKKMSI07}, we have given a lattice version of the decomposition (\ref{decomp1}) and (\ref{V-X}) for $SU(2)$ Yang-Mills theory.  
We have shown that the gauge-invariant magnetic monopole can be constructed in terms of $\mathscr{V}_\mu$ alone in such a way that the magnetic charge is integer-valued and subject to the Dirac quantization condition and that the magnetic contribution to the string tension defined from the variable $\mathscr{V}_\mu$ reproduces almost all the string tension (90\% of the full string tension) through a non-Abelian Stokes theorem for the  Wilson loop operator \cite{DP89,KondoIV}, yielding the magnetic monopole dominance \cite{IKKMSS06}.  Moreover, we have shown \cite{SKKMSI07} that   $\mathscr{V}_\mu(x)$ part is dominant in the infrared region, while  the remaining degrees of freedom represented by $\mathscr{X}_\mu(x)$ are suppressed by acquiring the large mass $M_X \cong 1.2 \sim 1.3$GeV, yielding the infrared ``Abelian" dominance \cite{tHooft81,EI82}.   
These are the gauge-invariant confirmations for remarkable results \cite{SY90,SNW94,AS99} which were first observed in the maximal Abelian gauge (MAG) \cite{KLSW87}, although the MAG breaks the original gauge group $SU(2)$ into U(1) explicitly. 

Now we turn our attention to the $SU(N)$ case. 
The issue of generating magnetic monopoles in Yang-Mills theory has been investigated so far   under the MAG which breaks the original gauge group $SU(N)$ into the maximal torus subgroup $H=U(1)^{N-1}$. Therefore, MAG yields $(N-1)$ types of magnetic monopole, in agreement with the observation due to the Homotopy group:
$\pi_2(SU(N)/U(1)^{N-1})=\pi_1(U(1)^{N-1})=\mathbb{Z}+\cdots+\mathbb{Z}$.
In the continuum formulation,  a possible extension of the CFNS decomposition from $SU(2)$ to $SU(N)$ group was worked out already in \cite{Cho80c,FN99a} where $N-1$ color fields $\bm{n}^j(x)$ ($j=1, \cdots, N-1)$ were introduced where $N-1$ is the rank of $SU(N)$, i.e., the dimension of the maximal torus subgroup $H=U(1)^{N-1}$. 
For $SU(N)$ Yang-Mills theory, therefore, it tends to assume that magnetic monopoles of $(N-1)$ types are necessary to cause the dual Meissner effect for realizing quark confinement. 
To the best of our knowledge, however, it is not yet confirmed  whether or not  $(N-1)$ type of magnetic monopole are necessary to achieve confinement in $SU(N)$ Yang-Mills theory.  
Rather, we have a conjecture that \textit{a single type of magnetic monopole is sufficient to achieve quark confinement even in $SU(N)$ Yang-Mills theory}, once it is defined in a gauge-invariant way.  In fact, this scenario was originally proposed in \cite{Kondo99Lattice99,KT99} based on a non-Abelian Stokes theorem for Wilson loop operator \cite{KT99,DP01}.   However, at that time, this idea was not substantiated because there did not exist the relevant lattice formulation which enables one to prove or disprove this conjecture. 

From this perspective, we consider new lattice descriptions of $SU(N)$ Yang-Mills theory in this Letter.  We must emphasize that possible ways of extending the $SU(2)$ machinery to $SU(N)$ case is not unique and there are several options.  
For the SU(3) gauge group, indeed, we have two options, maximal or minimal, although  there occur more involved  intermediate cases other than the maximal and minimal cases  for $N \ge 4$ in $SU(N)$ Yang-Mills theory \cite{KSM08}. 
In the maximal case, we can introduce $(N-1)$ unit vector fields $\bm{n}^j(x)$ which enable us to define $(N-1)$ types of magnetic monopoles.  This option corresponds to a straightforward extension of the $SU(2)$ lattice formulation given in previous papers \cite{KKMSSI05,IKKMSS06,SKKMSI07}.  Then the naive continuum limit of the maximal case on a lattice reduces to the $SU(N)$ CFNS decomposition in the continuum formulation \cite{Cho80c,FN99a}. 
Our gauge-invariant formulation in this option can reproduce the conventional results based on MAG by choosing the specific gauge, in which all the color fields are fixed to be the constant Cartan subalgebra on the whole lattice points, $\bm{n}^j(x) \rightarrow H^j$. 
In the minimal case, on the other hand, we introduce a single color field  yielding a single magnetic monopole even for $SU(N)$ Yang-Mills theory, as suggested from 
$\pi_2(SU(N)/U(N-1))=\pi_1(U(N-1))=\mathbb{Z}$.
Such a possibility seems to be overlooked in the previous works, although the continuum version of the minimal case was proposed and examined recently by three of us in a separate paper \cite{KSM08}. 
The lattice version given in this Letter is constructed so  as to agree with the continuum formulation \cite{KSM08} in the naive continuum limit. 
For this machinery to work, we give a prescription for constructing such a color field from the original Yang-Mills theory along the line shown in the continuum version.



\section{Maximal case}

For the Yang-Mills gauge theory with a gauge group $G=SU(N)$, it is convenient
\footnote{
But it is not essential to introduce $r$ fields $ {\bf n}^j(x)$, since it is enough to introduce a single color field $\bm{n}(x)$, see \cite{KSM08}.
}
 to introduce a set of  $(N^2-1)$-dimensional unit vector fields $ {\bf n}^j(x)$ 
($j=1, \cdots, r$) with the components $n_j^A(x)$, i.e., 
$
 {\bf n}^j(x) \cdot  {\bf n}^j(x) := n_j^A(x) n_j^A(x) = 1
$ 
($ A =1, 2, \dots, {\rm dim}G=N^2-1 $) 
where $r:={\rm rank}G=N-1$ is the rank  of the gauge group $G=SU(N)$.
We omit the summation symbol for $A$ in what follows. 
The   $\bm n^j(x)$ fields having the value in the Lie algebra $\mathscr{G}$ are constructed according to 
\begin{equation}
 \bm n^j(x) = n_j^A(x) T^A  = U^\dagger(x) H_j U(x) 
 , \quad U(x) \in G
 ,
\end{equation}
where $H_j$ are generators in the Cartan subalgebra in the generators $T^A$ of the Lie algebra $\mathscr{G}=su(N)$ of  $G=SU(N)$ and $U(x)$ is a group element of $G=SU(N)$. 
We adopt  the normalization ${\rm tr}(T_A T_B)= \frac12 \delta_{AB}$.

\subsection{Continuum: maximal case}

In the maximal case,  it has been shown \cite{KSM08} that the $su(N)$ Lie algebra valued Yang-Mills field  $\mathscr{A}_\mu(x)$ is decomposed into two parts:
\begin{align}
  \mathscr{A}_\mu(x) = \mathscr{V}_\mu(x) + \mathscr{X}_\mu(x) ,
\end{align}
 such  that all fields $\bm{n}^j(x)$ are covariantly constant in the background  field $\mathscr{V}_\mu(x)$:
\begin{subequations}
\begin{align}
 0 = \mathscr{D}_\mu[\mathscr{V}] \bm{n}^j(x) 
:= \partial_\mu \bm{n}^j(x) -i g [\mathscr{V}_\mu(x) , \bm{n}^j(x) ] 
\quad (j=1,2, \cdots, r) 
 ,
 \label{defVL}
\end{align}
and that the remaining field $\mathscr{X}_\mu(x)$ is orthogonal to all fields ${\bm n}^j(x)$:
\begin{align}
  0 = (\mathscr{X}_\mu(x), \bm{n}^j(x))   := 2{\rm tr}(\mathscr{X}_\mu(x) \bm{n}^j(x))  :=  \mathscr{X}_\mu^A(x) n_j^A(x)  
\quad (j=1,2, \cdots, r) 
\label{defXL}
 .
\end{align}
\end{subequations}
Both  $\mathscr{A}_\mu(x)$ and ${\bm n}^j(x)$ are Hermitian fields. This is also the case for $ \mathscr{V}_\mu(x)$ and $\mathscr{X}_\mu(x)$.

By solving the defining equation (\ref{defVL}) and (\ref{defXL}), $\mathscr{V}_\mu(x)$ and $\mathscr{X}_\mu(x)$ have been expressed in terms of $\mathscr{A}_\mu(x)$ and $\bm{n}^j(x)$ in the form:
\footnote{
In what follows, we use the script e.g., $\mathscr{A}_\mu$, $\bm{n}^j$ to express the $su(N)$ Lie algebra valued field and the boldface to express the vector field e.g., ${\bf n}^j$.
}
\begin{subequations}
\begin{align}
  \mathscr{V}_\mu(x) 
  =& \sum_{j=1}^{r}  (\mathscr{A}_\mu(x),{\bm n}^j(x)) {\bm n}^j(x)   +i g^{-1} \sum_{j=1}^{r} [ {\bm n}^j(x) , \partial_\mu {\bm n}^j(x) ] ,
  \label{Vdef}
\\
 \mathscr{X}_\mu(x) =& -ig^{-1} \sum_{j=1}^{r}  [\bm{n}^j(x), \mathscr{D}_\mu[\mathscr{A}]\bm{n}^j(x) ]
 .
 \label{Xdef}
\end{align}
\end{subequations}
In the maximal case, all $\bm{n}^j(x)$ fields are constructed from a single color field $\bm{n}(x)$. 
 Therefore, we have a  change of variables from $\mathscr{A}_\mu(x)$ to  ($\bm{n}(x)$, $c_\mu^j(x)$,  $\mathscr{X}_\mu(x)$) with $c_\mu^j(x):=(\mathscr{A}_\mu(x), \bm{n}^j(x))$, once    $\bm{n}(x)$ is given as a functional of $\mathscr{A}_\mu(x)$, as shown in \cite{KSM08}.

\subsection{Lattice: maximal case}

We try to decompose the $SU(N)$ link variable $U_{x,\mu}$ on a lattice into the product of two $SU(N)$ variables $X_{x,\mu}$ and $V_{x,\mu}$ on the same lattice:
\begin{equation}
 U_{x,\mu} = X_{x,\mu} V_{x,\mu} \in SU(N) ,
 \quad X_{x,\mu}, V_{x,\mu} \in SU(N)
  .
  \label{decomp-1}
\end{equation}
The link variable $U_{x,\mu}$ obeys the well-known lattice gauge transformation:
\begin{equation}
  U_{x,\mu}  \rightarrow \Omega_{x} U_{x,\mu} \Omega_{x+\mu}^\dagger = U_{x,\mu}^\prime
  , \quad \Omega_{x} \in SU(N)
  \label{U-transf}
 .
\end{equation}
We require that $V_{x,\mu}$ transforms like a usual gauge variable as 
\begin{subequations}
\begin{equation}
  V_{x,\mu}  \rightarrow \Omega_{x} V_{x,\mu} \Omega_{x+\mu}^\dagger = V_{x,\mu}^\prime 
  , \quad \Omega_{x} \in SU(N)
   .
   \label{V-transf}
\end{equation}
Consequently, $X_{x,\mu}$ must transform like an adjoint matter field:
\footnote{
It is possible to consider another decomposition of the form:
$
 U_{x,\mu} = V_{x,\mu} X_{x,\mu} 
  .
  \label{decomp-2}
$
Then  $X_{x,\mu}$ must have the gauge transformation:
$
  X_{x,\mu} (=  V_{x,\mu}^\dagger U_{x,\mu} )  \rightarrow \Omega_{x+\mu} X_{x,\mu} \Omega_{x+\mu}^\dagger  
  .
$
}
\begin{equation}
  X_{x,\mu} (=  U_{x,\mu}V_{x,\mu}^\dagger)
 \rightarrow \Omega_{x} X_{x,\mu} \Omega_{x}^\dagger = X_{x,\mu}^\prime
  , \quad \Omega_{x} \in SU(N)
  .
  \label{X-transf}
\end{equation}
At the same time, we introduce Hermitian variables ${\bm n}_{x}^j$  $(j=1, \cdots, r=N-1)$ on   sites of the lattice, $({\bm n}^j_{x})^\dagger={\bm n}^j_{x}$.  
Suppose that each site variable ${\bm n}^j_{x}$ transforms under the gauge transformation according to the adjoint representation as (\ref{n-ctransf}) suggests:
\begin{align}
  {\bm n}_{x}^j \rightarrow \Omega_{x} {\bm n}_{x}^j \Omega_{x}^\dagger = {\bm n}_{x}^j{}' , \quad \Omega_{x} \in SU(N)
\label{n-transf}
 .
\end{align}
\end{subequations}
Once such $SU(N)$ lattice variables $V_{x,\mu}$ and $X_{x,\mu}$ are constructed together with ${\bm n}^j_{x}$, they are related to the Lie algebra valued fields $\mathscr{V}_\mu$ and $\mathscr{X}_\mu$ in the form:
\begin{align}
  V_{x,\mu} = \exp [-i\epsilon g \mathscr{V}_\mu(x+\epsilon\mu/2)] 
  , \quad
  X_{x,\mu} = \exp [-i\epsilon g \mathscr{X}_\mu(x)]
 ,
\end{align}
just as the original link variable $U_{x,\mu}$ is related to the gauge potential $\mathscr{A}_\mu(x)$: 
\begin{align}
U_{x,\mu} = \exp [ -i \epsilon g \mathscr{A}_\mu(x+\epsilon\mu/2)]
 , 
\label{def-U}
\end{align}
where $\epsilon$ is the lattice spacing.
The transformation property (\ref{V-transf}) for the group element $V_{x,\mu}$ in $SU(N)$ is required from   (\ref{V-ctransf}) of the $su(N)$ valued Hermitian  variable $\mathscr{V}_\mu(x)$, 
which is the background field $\mathscr{V}_\mu(x)$ to be identified with the continuum  variable (\ref{Vdef}) in the continuum limit $\epsilon \rightarrow 0$. 
Similarly, the transformation (\ref{X-transf}) for the other variable  $X_{x,\mu}$ is  consistent with  (\ref{X-ctransf})  of the $su(N)$ valued Hermitian  variable $\mathscr{X}_\mu(x)$ to be identified with the continuum  variable (\ref{Xdef}) in the continuum limit. 
For details of the continuum formulations, see \cite{KSM08} for $SU(N)$ and especially \cite{KMS06} for $SU(2)$ case.
We first try to impose the following conditions as the lattice versions  of the defining equations (\ref{defVL}) and (\ref{defXL}) for $SU(N)$, which is a straightforward extension of the $SU(2)$ case \cite{IKKMSS06,SKKMSI07}. 
\begin{subequations}
\begin{align}
 & \text{(a)} \quad 
 \bm{n}_{x}^j V_{x,\mu}  = V_{x,\mu} \bm{n}_{x+\mu}^j   
  ,
 \label{Lcc}
\\
 & \text{(b)} \quad
 {\rm tr}(\bm{n}_{x}^j U_{x,\mu} V_{x,\mu}^\dagger) 
    =  0 = {\rm tr}(\bm{n}_{x}^j V_{x,\mu} U_{x,\mu}^\dagger) 
 .
  \label{cond2m}
\end{align}
\end{subequations}
The last one is equivalent to 
$
 {\rm tr}(\bm{n}_{x}^j X_{x,\mu} ) 
  = 0 = {\rm tr}(\bm{n}_{x}^j X_{x,\mu}^\dagger) 
 .
$
Both defining equations must be solved to determine new variables $V_{x,\mu}$ and $X_{x,\mu}$ for a given set of $\bm{n}_{x}$ and $U_{x,\mu}$.  
The first defining equation (\ref{Lcc}) for the link variable $V_{x,\mu}$  is form-invariant under the gauge transformation for new variables, i.e.,
$ \bm{n}_{x}^{j \prime} V_{x,\mu}^\prime  = V_{x,\mu}^\prime \bm{n}_{x+\mu}^{j \prime}
$. 
This is also the case for the second defining equation (\ref{cond2m}). 
The  (\ref{defVL}) comes from the lattice covariant derivative in the adjoint representation under the background ${\bf V}_\mu(x)$ \cite{SKKMSI07}:
\begin{align}
 D_\mu^{(\epsilon)}[V] \bm{n}_{x}^j := \epsilon^{-1}[V_{x,\mu} \bm{n}_{x+\mu}^j - \bm{n}_{x}^j V_{x,\mu}]
  .  
 \label{cderivative}
\end{align}
The  (\ref{cond2m}) is suggested from a lattice version of the orthogonality equation (\ref{defXL}): 
${\rm tr}({\bm n}_{x}^j \mathscr{X}_\mu(x))=0$ or
\begin{equation}
 {\rm tr}({\bm n}^j_{x} \exp \{-i\epsilon g \mathscr{X}_\mu(x)\})
  =  {\rm tr}({\bm n}^j_{x}  \{ {\bf 1}-i\epsilon g \mathscr{X}_\mu(x) \} ) + O(\epsilon^2) = 0 + O(\epsilon^2) .
  \label{cond2}
\end{equation}
This implies that the trace vanishes  up to   $O(\epsilon)$. 
Remembering  the relation $\mathscr{X}_\mu(x)=\mathscr{A}_\mu(x)-\mathscr{V}_\mu(x)$, we can rewrite   (\ref{cond2}) into (\ref{cond2m}) in terms of ${\bm n}_{x}$ and $U_{x,\mu}$.

\subsection{How to solve the lattice defining equations?}

First, we consider how the defining equation (\ref{Lcc}) is solved to express the link variable $V_{x,\mu}$ in terms of the site variable ${\bm n}^j_{x}$ and the original link variable $U_{x,\mu}$, 
just as the continuum variable $\mathscr{V}_\mu(x)$ is expressed in terms of ${\bm n}^j(x)$ and $\mathscr{A}_\mu(x)$ in (\ref{Vdef}). 
In order to solve the matrix equation (\ref{Lcc}), we adopt an ansatz: 
\begin{align}
 \tilde V_{x,\mu}
 = c U_{x,\mu} + \sum_{j=1}^{r} \alpha_j {\bm n}^j_{x} U_{x,\mu} + \sum_{j=1}^{r} \beta_j U_{x,\mu} {\bm n}^j_{x+\mu} 
 + \sum_{j=1}^{r} \sum_{k=1}^{r} \gamma_{jk} {\bm n}^j_{x} U_{x,\mu} {\bm n}^k_{x+\mu}
  ,
 \label{ansatz}
\end{align}
where $c, \alpha_j, \beta_k, \gamma_{jk}$ ($j,k=1,\cdots,r$) are complex numbers to be determined.
In fact, $\tilde V_{x,\mu}$ has the transformation property (\ref{V-transf}) required for the link variable $V_{x,\mu}$, provided that the original link variable $U_{x,\mu}$ and the color field ${\bm n}^j_x$ transform  according to (\ref{U-transf}) and (\ref{n-transf}), respectively.
At first glance, this ansatz is rather specific in that it is linear in the link variable $U_{x,\mu}$ and contains the color field ${\bm n}^j_x$ up to quadratic in all ${\bm n}^j$. 
However, this is the most general form within the general linear matrix $GL(N, \mathbb{C})$, since  the ansatz involves $1+r+r+r^2=N^2$ complex parameters which is equal to the dimension of a matrix in $GL(N, \mathbb{C})$. 
Moreover, the first equation (\ref{Lcc}) is linear in  $V_{x,\mu}$ and the normalization of $\tilde{V}_{x,\mu}$ cannot be determined by this equation alone.  In what follows, therefore, we set $c=1$ hereafter without loss of generality. 

We show that the solution of the defining equation based on the ansatz (\ref{ansatz}) for $\tilde V_{x,\mu}$ is sufficient to obtain the special unitary link variable $V_{x,\mu}$ satisfying the defining equations.
First, we show that the unitary link variable $\underbar{V}_{x,\mu}$ can be constructed from $\tilde V_{x,\mu}$ based on the {\it polar decomposition theorem}, 
\footnote{
This is a complex version of the polar decomposition theorem:  Any real regular (invertible) matrix $M$ is uniquely expressed by an orthogonal matrix $U$ and a positive symmetric matrix $P$ as
$
 M =PU = UQ, \quad P = \sqrt{MM^T}, \quad Q = \sqrt{M^T M}
    .
$
}
although the unitarity of the link variable $\tilde V_{x,\mu}$ of the ansatz (\ref{ansatz}) is not guaranteed. 

\noindent
{\bf Theorem} (Polar decomposition theorem)
{\it Any complex regular matrix $M$ is uniquely decomposed into a unitary matrix $U$ and a positive-definite Hermitian matrix $P$ as
 \footnote{
Any complex regular matrix $M$ is also decomposed into another  positive-definite Hermitian matrix $Q$ as
$
 M = UQ
$, 
$Q = \sqrt{M^\dagger M}
$. 
}
\begin{equation}
 M = PU  , \quad P = \sqrt{MM^\dagger }
  .
\end{equation}
}
Thus the unitary link variable $\underbar{V}_{x,\mu} \in U(N)$ is obtained in terms of $\tilde{V}_{x,\mu}$ as
\footnote{
Here
$\underbar{V}_{x,\mu} = \tilde V_{x,\mu} Q_{x,\mu}^{-1}$ must be used for another case 
$
 U_{x,\mu} = V_{x,\mu} X_{x,\mu} 
$
where 
$
 Q_{x,\mu} = \sqrt{\tilde V_{x,\mu}^\dagger  \tilde V_{x,\mu}} 
    .
$
}
\begin{equation}
\underbar{V}_{x,\mu} 
= P_{x,\mu}^{-1} \tilde V_{x,\mu}  
 ,
\quad 
 P_{x,\mu} = \sqrt{\tilde V_{x,\mu}  \tilde V_{x,\mu}^\dagger}  
  .
\label{polar}
\end{equation}
Moreover,  the special unitary link variable $V_{x,\mu} \in SU(N)$ is obtained by the normalization which guarantees $\det V_{x,\mu}=1$:
\footnote{
For $SU(2)$, in particular, $ \tilde V_{x,\mu}  \tilde V_{x,\mu}^\dagger$ (or $ \tilde V_{x,\mu}^\dagger  \tilde V_{x,\mu}$) is proportional to the unit matrix and  the operation $P^{-1}$ (or $Q^{-1}$) is  equal to multiplying a normalization factor 
$1/\sqrt{\frac12 {\rm tr}( \tilde V_{x,\mu}^\dagger  \tilde V_{x,\mu})}=1/\sqrt{\frac12 {\rm tr}( \tilde V_{x,\mu} \tilde V_{x,\mu}^\dagger)}$  \cite{IKKMSS06}.
This is not the case for $SU(N)$, $N \ge 3$.
}
\begin{equation}
 V_{x,\mu} = \underbar{V}_{x,\mu}/(\det \underbar{V}_{x,\mu})^{1/N}
  .
  \label{speciality}
\end{equation}

Next, we show that the first defining equation (\ref{Lcc}) is automatically satisfied for $V_{x,\mu}$ which are made unitary by (\ref{polar}), if   $\tilde V_{x,\mu}$ satisfy the same equation (a'):
\begin{align}
\text{(a')}  \quad\quad
  \bm{n}_{x}^j \tilde V_{x,\mu}  
= \tilde V_{x,\mu} \bm{n}_{x+\mu}^j
 .
 \label{Lcc2}
\end{align}
  If $\tilde V_{x,\mu}$ satisfies the defining equation (a'), $\tilde V_{x,\mu} \tilde V_{x,\mu}^\dagger$ commutes with $\bm{n}_{x}^j$, since 
\begin{align}
 {\bf n}_{x}^j \tilde V_{x,\mu} \tilde V_{x,\mu}^\dagger  
= \tilde V_{x,\mu} {\bf n}_{x+\mu}^j  \tilde V_{x,\mu}^\dagger
=  \tilde V_{x,\mu}  \tilde V_{x,\mu}^\dagger {\bf n}_{x}^j
 \Longleftrightarrow  [ \tilde V_{x,\mu} \tilde V_{x,\mu}^\dagger , {\bf n}_{x}^j ] = 0
\label{VVd}
  ,
\end{align}
while $\tilde V_{x,\mu}^\dagger \tilde V_{x,\mu}$ commutes with ${\bf n}_{x+\mu}^j$:
$
   \tilde V_{x,\mu}^\dagger  \tilde V_{x,\mu} {\bf n}_{x+\mu}^j
= \tilde V_{x,\mu}^\dagger {\bf n}_{x}^j \tilde V_{x,\mu}   
=   {\bf n}_{x+\mu}^j \tilde V_{x,\mu}^\dagger \tilde V_{x,\mu}  
 \Longleftrightarrow  [ \tilde V_{x,\mu}^\dagger  \tilde V_{x,\mu} , {\bf n}_{x+\mu}^j ] = 0
  .
$
It is also possible to prove (see Appendix~\ref{appendix:spectral-resolution}) that  ${\bf n}_{x}^j$   commutes  with the square root $P_{x,\mu}$ and its inverse   square root $P_{x,\mu}^{-1}$,
\footnote{
Whereas 
$
  [ \tilde V_{x,\mu}^\dagger \tilde V_{x,\mu}  , {\bf n}_{x+\mu}^j ] = 0 \Longleftrightarrow
 [ \sqrt{\tilde V_{x,\mu}^\dagger \tilde V_{x,\mu} } , {\bf n}_{x+\mu}^j ] = 0
 \Longleftrightarrow
  [ Q_{x,\mu}^{-1} , {\bf n}_{x}^j ] = 0 
 .
$
} 
\begin{equation}
  [ \tilde V_{x,\mu} \tilde V_{x,\mu}^\dagger , {\bf n}_{x}^j ] = 0 
\Longleftrightarrow
 [ \sqrt{\tilde V_{x,\mu} \tilde V_{x,\mu}^\dagger} , {\bf n}_{x}^j ] = 0
 \Longleftrightarrow
  [ P_{x,\mu}^{-1} , {\bf n}_{x}^j ] = 0 
 .
\label{VdV2}
\end{equation}
To obtain  $V_{x,\mu}$ satisfying (a), therefore, it is sufficient to find a solution $\tilde V_{x,\mu}$ of   (a'), since 
(\ref{VdV2}) leads to 
\footnote{
Whereas 
$
 {\bf n}_{x}^j  \underbar{V}_{x,\mu} 
=  {\bf n}_{x}^j \tilde V_{x,\mu} Q_{x,\mu}^{-1}
= \tilde V_{x,\mu} {\bf n}_{x+\mu}^j  Q_{x,\mu}^{-1}
= \tilde V_{x,\mu}  Q_{x,\mu}^{-1} {\bf n}_{x+\mu}^j  
=  \underbar{V}_{x,\mu} {\bf n}_{x+\mu}^j  
 .
$
}
\begin{align}
 {\bf n}_{x}^j  \underbar{V}_{x,\mu} 
=&  {\bf n}_{x}^j  P_{x,\mu}^{-1} \tilde V_{x,\mu} 
=   P_{x,\mu}^{-1} {\bf n}_{x}^j  \tilde V_{x,\mu}  
=  P_{x,\mu}^{-1} \tilde V_{x,\mu}   {\bf n}_{x+\mu}^j  
=  \underbar{V}_{x,\mu} {\bf n}_{x+\mu}^j  
  .
\end{align}

For the second defining equation, we adopt the following  (b') instead of (b).
\begin{equation}
\text{(b')}  \quad \quad
 {\rm tr}(\bm{n}_{x} \tilde X_{x,\mu})=0, \quad
   \tilde  X_{x,\mu} = U_{x,\mu} \tilde V_{x,\mu}^{-1}
( \ne U_{x,\mu} \tilde V_{x,\mu}^\dagger ) 
  ,
  \label{def-2b}
\end{equation}
or more calculable form:
$\text{(b'')}$ 
$   {\rm tr}(\bm{n}_{x} \tilde X_{x,\mu}^{-1})=0, 
$
$\tilde X_{x,\mu}^{-1} = \tilde  V_{x,\mu} U_{x,\mu}^{-1} = \tilde  V_{x,\mu} U_{x,\mu}^\dagger 
  .
$
Although (b') is written for $\tilde X_{x,\mu}$ and it does not guarantee the validity of   (b) for $X_{x,\mu}$,  two conditions (b) and (b') have no difference in the level of the Lie algebra which is relevant in the continuum limit and that (b') yield (\ref{defXL}) in the continuum limit.

For $SU(N)$ ($N \ge 3$),  thus, the decomposition of the original link variable is achieved, once we know $\tilde V_{x,\mu}$ satisfying the defining equation (a'), (b'): 
\footnote{
The identification
$X_{x,\mu} 
=  Q_{x,\mu} \tilde X_{x,\mu}$ must be used for another case 
$
 U_{x,\mu} = V_{x,\mu} X_{x,\mu} 
$.
}
\begin{subequations}
\begin{align}
SU(N)  \ni U_{x,\mu} 
=& X_{x,\mu} V_{x,\mu} 
=  \tilde X_{x,\mu} \tilde V_{x,\mu}    ,
  \quad \tilde X_{x,\mu}, \tilde V_{x,\mu} \notin SU(N)
  ,
  \label{decomp-f}
\\
  V_{x,\mu}  =&  P_{x,\mu}^{-1} \tilde V_{x,\mu}/(\det (P_{x,\mu}^{-1} \tilde V_{x,\mu}))^{1/N}  \in SU(N)
    , \quad 
    P_{x,\mu} := \sqrt{\tilde V_{x,\mu}  \tilde V_{x,\mu}^\dagger} 
\label{polar11}
\\
   X_{x,\mu} 
=&  U_{x,\mu} \tilde V_{x,\mu}^{-1} P_{x,\mu} (\det (P_{x,\mu}^{-1} \tilde V_{x,\mu}))^{1/N} \in SU(N)
  .
\label{polar-X}
\end{align}
\end{subequations}
Thus, solving the defining equations (a'):(\ref{Lcc2}) and (b'):(\ref{def-2b}) under the ansatz (\ref{ansatz}) is sufficient to achieve the desired decomposition through the identification (\ref{decomp-f}).

\subsection{Solving the lattice defining equations}

In order to see when the first defining equation (a') meets, we calculate 
\begin{align}
 {\bm n}^\ell_{x} {\tilde V}_{x,\mu}
 =& {\bm n}^\ell_{x} U_{x,\mu} + \sum_{j=1}^{r} \alpha_j {\bm n}^\ell_{x} {\bm n}^j_{x} U_{x,\mu} + \sum_{j=1}^{r} \beta_j {\bm n}^\ell_{x} U_{x,\mu} {\bm n}^j_{x+\mu} 
 + \sum_{j=1}^{r} \sum_{k=1}^{r} \gamma_{jk} {\bm n}^\ell_{x} {\bm n}^j_{x} U_{x,\mu} {\bm n}^k_{x+\mu} 
  \nonumber\\
 =&  
  \alpha_\ell 
  \frac{1}{2N}  U_{x,\mu} 
+ \left[ {\bm n}^\ell_{x} U_{x,\mu}
+ \sum_{j=1}^{r} \alpha_j  \frac{1}{2} (\bm n^\ell * \bm n^j)_{x}  U_{x,\mu} \right] 
 +  \sum_{k=1}^{r} \gamma_{\ell k}
 \frac{1}{2N} U_{x,\mu} {\bm n}^k_{x+\mu}
 \nonumber\\
 &+ \left[ \sum_{j=1}^{r} \beta_j {\bm n}^\ell_{x} U_{x,\mu} {\bm n}^j_{x+\mu} 
  + \sum_{j=1}^{r} \sum_{k=1}^{r} \gamma_{jk}  \frac{1}{2} (\bm n^\ell * \bm n^j)_{x}  
U_{x,\mu} {\bm n}^k_{x+\mu} \right] 
 .
 \label{ansatz1}
\end{align}
Here we have used  
$ T^A T^B 
=  \frac12 [ T^A , T^B  ] +  \frac12 \{ T^A , T^B  \} 
=  \frac{i}{2} f^{ABC} T^C + \frac{1}{2N} \delta^{AB} {\bf 1} + \frac{1}{2} d^{ABC} T^C 
$
and 
$
 \bm n^j \bm n^k 
 =  \frac12 [ \bm n^j ,\bm n^k ] +  \frac12 \{ \bm n^j ,\bm n^k \} 
= \frac{1}{2N} \delta_{jk} {\bf 1} + \frac{1}{2} (\bm n^j * \bm n^k) 
  ,
$
since $[\bm n^j ,\bm n^k ]=0$, 
where we have defined  
$
\bm n^j * \bm n^k 
:= ({\bf n}^j * {\bf n}^k)^C T^C 
:= d^{ABC} n_j^A n_k^B T^C 
 .
$
On the other hand, we have
\begin{align}
 {\tilde V}_{x,\mu} {\bm n}^\ell_{x+\mu} 
 =& U_{x,\mu} {\bm n}^\ell_{x+\mu} + \sum_{j=1}^{r} \alpha_j {\bm n}^j_{x} U_{x,\mu} {\bm n}^\ell_{x+\mu} + \sum_{j=1}^{r} \beta_j U_{x,\mu} {\bm n}^j_{x+\mu} {\bm n}^\ell_{x+\mu} 
 + \sum_{j=1}^{r} \sum_{k=1}^{r} \gamma_{jk} {\bm n}^j_{x} U_{x,\mu} {\bm n}^k_{x+\mu} {\bm n}^\ell_{x+\mu}
 \nonumber\\
 =& 
 \frac{1}{2N}  \beta_\ell U_{x,\mu}  
+ \sum_{j=1}^{r} 
 \frac{1}{2N} \gamma_{j\ell} {\bm n}^j_{x} U_{x,\mu} 
+ \left[ U_{x,\mu} {\bm n}^\ell_{x+\mu} 
+  \sum_{j=1}^{r} \beta_j \frac{1}{2} U_{x,\mu} (\bm n^j * \bm n^\ell)_{x+\mu}   \right] 
 \nonumber\\
 & + \left[ \sum_{j=1}^{r} \alpha_j {\bm n}^j_{x} U_{x,\mu} {\bm n}^\ell_{x+\mu} 
+ \sum_{j=1}^{r} \sum_{k=1}^{r} \gamma_{jk} \frac{1}{2}  {\bm n}^j_{x} U_{x,\mu}  (\bm n_k * \bm n^\ell)_{x+\mu}  \right] 
  .
 \label{ansatz2}
\end{align}
The four terms in the right-hand sides of (\ref{ansatz1}) and (\ref{ansatz2}) are independent.  In order to fulfill (a'), therefore, the following conditions must be satisfied simultaneously:
\begin{subequations}
\begin{align}
  \text{(i)} \quad &
\alpha_\ell = \beta_\ell
  ,
\\
  \text{(ii)} \quad &
 {\bm n}^\ell_{x} U_{x,\mu}
+ \sum_{j=1}^{r} \alpha_j  \frac{1}{2} (\bm n^\ell * \bm n^j)_{x}  U_{x,\mu} 
=\sum_{j=1}^{r} 
 \frac{1}{2N} \gamma_{j\ell} {\bm n}^j_{x} U_{x,\mu} 
 ,
\\
  \text{(iii)} \quad &
 \sum_{k=1}^{r} \gamma_{\ell k}
 \frac{1}{2N} U_{x,\mu} {\bm n}^k_{x+\mu}
= U_{x,\mu} {\bm n}^\ell_{x+\mu}
+ \sum_{j=1}^{r} \beta_j \frac{1}{2}  U_{x,\mu} (\bm n^j * \bm n^\ell)_{x+\mu}  
 ,
\\
  \text{(iv)} \quad &
  \sum_{j=1}^{r} \beta_j {\bm n}^\ell_{x} U_{x,\mu} {\bm n}^j_{x+\mu} 
  + \sum_{j=1}^{r} \sum_{k=1}^{r} \gamma_{jk}  \frac{1}{2} (\bm n^\ell * \bm n^j)_{x}  
U_{x,\mu} {\bm n}^k_{x+\mu} 
\nonumber\\
&= \sum_{j=1}^{r} \alpha_j {\bm n}^j_{x} U_{x,\mu} {\bm n}^\ell_{x+\mu} 
+ \sum_{j=1}^{r} \sum_{k=1}^{r} \gamma_{jk}\frac{1}{2}  {\bm n}^j_{x} U_{x,\mu}  (\bm n^k * \bm n^\ell)_{x+\mu} 
 ,
\end{align}
\end{subequations}
where the symmetric product of two $\bm n^j$ fields is written as a linear combination of some of $\bm n^j$ fields. 
For $SU(N)$, the relevant information is given in \cite{KSM08}.

For $SU(2)$, the symmetric product is identically zero: $d^{ABC}=0$:
\begin{align}
 \bm n  \bm n   = (1/4) {\bf 1}  , \quad
 \bm n * \bm n  = 0
  .
\end{align} 
Therefore, we immediately find the solution:
$\alpha_1=\beta_1$, $\gamma_{11}=4$,
and 
\begin{align}
 {\tilde V}_{x,\mu}
 = U_{x,\mu} +  \alpha_1  ({\bm n}_{x} U_{x,\mu} +   U_{x,\mu} {\bm n}_{x+\mu} )
 +  4{\bm n}_{x} U_{x,\mu} {\bm n}_{x+\mu}
  .
 \label{SU2}
\end{align}
Therefore, one parameter $\alpha_1$ is undetermined. 
This result reproduces the previous one \cite{IKKMSS06}. (This result for the value $\gamma_{11}=4$ is apparently different from the previous one due to the fact that we adopt the different normalization for the color field.)

For $SU(3)$, we have three relations for two  fields $\bm{n}^1$ and $\bm{n}^2$, 
\begin{align}
{\bf n}^1 * {\bf n}^1  =& (1/\sqrt{3}) {\bf n}^2 
 ,
\nonumber\\
{\bf n}^1 * {\bf n}^2 =& {\bf n}^2 * {\bf n}^1 = (1/\sqrt{3}){\bf n}^1 
 ,
\nonumber\\
{\bf n}^2 * {\bf n}^2 =& -(1/\sqrt{3}) {\bf n}^2 
\label{nSU3}
 .
\end{align}
For $SU(3)$, the ansatz reads
\begin{align}
{\tilde V}_{x,\mu} =& {\tilde V}_{x,\mu}^{(+)} + {\tilde V}_{x,\mu}^{(-)} ,
 \nonumber\\
{\tilde V}_{x,\mu}^{(+)}
 &=U_{x,\mu}
   +\alpha_2 {\bm n^2_x}U_{x,\mu}
   +\beta_2 U_{x,\mu}{\bm n^2_{x+\mu}}
   +\gamma_{11}{\bm n^1_x}U_{x,\mu}{\bm n^1_{x+\mu}}
   +\gamma_{22} {\bm n^2_x}U_{x,\mu}{\bm n^2_{x+\mu}}
 ,
 \nonumber\\
{\tilde V}_{x,\mu}^{(-)}
 &=\alpha_1 {\bm n^1_x}U_{x,\mu}
   +\beta_1 U_{x,\mu}{\bm n^1_{x+\mu}}
   +\gamma_{12} {\bm n^1_x}U_{x,\mu}{\bm n^2_{x+\mu}}
   +\gamma_{21} {\bm n^2_x}U_{x,\mu}{\bm n^1_{x+\mu}} 
    ,
\end{align}
where we have separated ${\tilde V}_{x,\mu}$ into two parts according to the parity under the global transformation $\bm n^1_x \rightarrow-\bm n^1_x$, under which $\bm n^2_x$ does not change.

The requirement (ii) reads 
for $\ell=1$, 
\begin{align}
0 =&  {\bm n}^1_{x} U_{x,\mu}
+  \alpha_1  \frac{1}{2} (\bm n_1 * \bm n_1)_{x}  U_{x,\mu} 
+  \alpha_2  \frac{1}{2} (\bm n_1 * \bm n_2)_{x}  U_{x,\mu} 
- \frac{1}{6} \gamma_{11} {\bm n}^1_{x} U_{x,\mu} 
 - \frac{1}{6} \gamma_{21} {\bm n}^2_{x} U_{x,\mu} 
 \nonumber\\
=&  {\bm n}^1_{x} U_{x,\mu}
+  \alpha_1  \frac{1}{2}  \frac{1}{\sqrt{3}} \bm n^2_{x}  U_{x,\mu} 
+  \alpha_2  \frac{1}{2} \frac{1}{\sqrt{3}} \bm n^1_{x} U_{x,\mu} 
- \frac{1}{6} \gamma_{11} {\bm n}^1_{x} U_{x,\mu} 
 - \frac{1}{6} \gamma_{21} {\bm n}^2_{x} U_{x,\mu} 
 ,
\end{align}
while for $\ell=2$, 
\begin{align}
0 =&  {\bm n}^2_{x} U_{x,\mu}
+  \alpha_1  \frac{1}{2} (\bm n_2 * \bm n_1)_{x}  U_{x,\mu} 
+  \alpha_2  \frac{1}{2} (\bm n_2 * \bm n_2)_{x}  U_{x,\mu} 
- \frac{1}{6} \gamma_{12} {\bm n}^1_{x} U_{x,\mu} 
 - \frac{1}{6} \gamma_{22} {\bm n}^2_{x} U_{x,\mu} 
 \nonumber\\
=&  {\bm n}^2_{x} U_{x,\mu}
+  \alpha_1  \frac{1}{2}  \frac{1}{\sqrt{3}} \bm n^1_{x}  U_{x,\mu} 
-  \alpha_2  \frac{1}{2} \frac{1}{\sqrt{3}} \bm n^2_{x} U_{x,\mu} 
- \frac{1}{6} \gamma_{12} {\bm n}^1_{x} U_{x,\mu} 
 - \frac{1}{6} \gamma_{22} {\bm n}^2_{x} U_{x,\mu} 
 .
\end{align}
Then we easily find 
\begin{align}
& \alpha_1=\beta_1, \quad \alpha_2=\beta_2,  
 \nonumber\\
 & \gamma_{11} = 6 +\sqrt{3} \alpha_2 , \quad \gamma_{21} = \sqrt{3} \alpha_1
 , \quad
 \gamma_{22} = 6 -\sqrt{3} \alpha_2 , \quad \gamma_{12} = \sqrt{3} \alpha_1
 .
 \label{SU3-parameter}
\end{align}
Thus the requirement (a') yields the result: 
\begin{align}
{\tilde V}_{x,\mu}^{(+)}
 &=U_{x,\mu}  +\alpha_2({\bm n^2_x}U_{x,\mu}+U_{x,\mu}{\bm n^2_{x+\mu}})
   \nonumber\\
 &   +(6+\sqrt3\alpha_2){\bm n^1_x}U_{x,\mu}{\bm n^1_{x+\mu}}
   +(6-\sqrt3\alpha_2){\bm n^2_x}U_{x,\mu}{\bm n^2_{x+\mu}}
 ,
 \nonumber\\
{\tilde V}_{x,\mu}^{(-)}
 &=\alpha_1 ({\bm n^1_x}U_{x,\mu}
   + U_{x,\mu}{\bm n^1_{x+\mu}}
   +\sqrt3 {\bm n^1_x}U_{x,\mu}{\bm n^2_{x+\mu}}
   +\sqrt3 {\bm n^2_x}U_{x,\mu}{\bm n^1_{x+\mu}}) 
 .
\label{SU3-1}
\end{align}
Therefore, two parameters $\alpha_1$ and $\alpha_2$ are still undetermined and should be fixed by imposing the requirement (b') or (b''), in the similar way to the $SU(2)$ case.

Next, we examine the second defining equation (b'). We rewrite the product 
\begin{align}
 {\bm n}^\ell_{x} {\tilde V}_{x,\mu}U_{x,\mu}^\dagger
 =&{\bm n}^\ell_{x} + \sum_{j=1}^{r} \alpha_j {\bm n}^\ell_{x}{\bm n}^j_{x}  + \sum_{j=1}^{r} \beta_j {\bm n}^\ell_{x} U_{x,\mu} {\bm n}^j_{x+\mu} U_{x,\mu}^\dagger
 + \sum_{j=1}^{r} \sum_{k=1}^{r} \gamma_{jk} {\bm n}^\ell_{x} {\bm n}^j_{x} U_{x,\mu} {\bm n}^k_{x+\mu}U_{x,\mu}^\dagger
\nonumber\\
 =&   \alpha_\ell   \frac{1}{2N} {\bf 1} + {\bm n}^\ell_{x} +  \sum_{j=1}^{r}  \alpha_j  \frac12 ({\bm n}^\ell*{\bm n}^j)_{x}   
+   \sum_{k=1}^{r} \gamma_{\ell k}   \frac{1}{2N}    U_{x,\mu} {\bm n}^k_{x+\mu}U_{x,\mu}^\dagger 
\nonumber\\
 & + \sum_{j=1}^{r} \beta_j {\bm n}^\ell_{x} U_{x,\mu} {\bm n}^j_{x+\mu} U_{x,\mu}^\dagger
+ \sum_{j=1}^{r} \sum_{k=1}^{r} \gamma_{jk} \frac12 ({\bm n}^\ell*{\bm n}^j)_{x} U_{x,\mu} {\bm n}^k_{x+\mu}U_{x,\mu}^\dagger
  .
\end{align}
The trace of ${\bm n}^\ell_{x} {\tilde V}_{x,\mu}U_{x,\mu}^\dagger$ reads 
\begin{align}
 {\rm tr}({\bm n}^\ell_{x} {\tilde V}_{x,\mu}U_{x,\mu}^\dagger)
 =&  
 \alpha_\ell   \frac{1}{2}
  + \sum_{j=1}^{r} \beta_j {\rm tr}({\bm n}^\ell_{x} U_{x,\mu} {\bm n}^j_{x+\mu} U_{x,\mu}^\dagger)
\nonumber\\&
 + \sum_{j=1}^{r} \sum_{k=1}^{r} \frac{1}{2} \gamma_{jk} {\rm tr}(({\bm n}^\ell_{x}*{\bm n}^j_{x}) U_{x,\mu} {\bm n}^k_{x+\mu}U_{x,\mu}^\dagger)
  ,
\end{align}
where we have used
${\rm tr}({\bm n}^\ell_{x})=0$ and ${\rm tr}({\bm n}^\ell_{x}*{\bm n}^j_{x})=0$.
The term ${\bm n}^k_{x} U_{x,\mu} {\bm n}^j_{x+\mu} U_{x,\mu}^\dagger$ has the expansion in the Lie algebra:
\begin{align}
& {\bm n}^k_{x} U_{x,\mu} {\bm n}^j_{x+\mu} U_{x,\mu}^\dagger 
\nonumber\\
&= {\bm n}^k_{x}[{\bf 1}-i\epsilon g\mathscr{A}_\mu(x)+ O((\epsilon^2)][{\bm n}^j_{x}+\epsilon \partial_\mu {\bm n}^j_{x}+ O((\epsilon^2)][1+i\epsilon g\mathscr{A}_\mu(x)+ O((\epsilon^2)]  
\nonumber\\
&= {\bm n}^k_{x} {\bm n}^j_{x} + i {\bm n}^k_{x} {\bm n}^j_{x} \epsilon g\mathscr{A}_\mu(x) + \epsilon {\bm n}^k_{x} \partial_\mu {\bm n}^j_{x} -i {\bm n}^k_{x} \epsilon g\mathscr{A}_\mu {\bm n}^j_{x} +  O((\epsilon^2) 
  .
\end{align}
Then its trace is given by 
\begin{align}
& {\rm tr}({\bm n}^k_{x} U_{x,\mu} {\bm n}^j_{x+\mu} U_{x,\mu}^\dagger)
\nonumber\\
&= {\rm tr}({\bm n}^k_{x} {\bm n}^j_{x}) + i {\rm tr}([{\bm n}^k_{x}, {\bm n}^j_{x}] \epsilon g\mathscr{A}_\mu(x)) + {\rm tr}(\epsilon {\bm n}^k_{x} \partial_\mu {\bm n}^j_{x})  +  O((\epsilon^2)  
\nonumber\\
&= \frac{1}{2} \delta_{kj}    +  O((\epsilon^2)  
\label{tr(nUnU)}
  ,
\end{align}
where we have used 
$[{\bm n}^k_{x}, {\bm n}^j_{x}]=0$ and ${\rm tr}({\bm n}^\ell_{x})=0$.
Taking into account this result, thus, we obtain another set of conditions:
\begin{align}
0   =  {\rm tr}({\bm n}^\ell_{x} {\tilde V}_{x,\mu}U_{x,\mu}^\dagger) =   \alpha_\ell   
+ \sum_{j=1}^{r} \sum_{k=1}^{r} \frac{1}{4} \gamma_{jk} 
 {\rm tr}(({\bm n}^\ell_{x}*{\bm n}^j_{x}) U_{x,\mu} {\bm n}^k_{x+\mu}U_{x,\mu}^\dagger)
  ,
  \label{nVU}
\end{align}
where the last trace is calculated from (\ref{tr(nUnU)}) after applying (\ref{nSU3})  to  $({\bm n}^\ell_{x}*{\bm n}^j_{x})$. 

For $SU(2)$, we immediately have
$
 \alpha_1 = 0
$
and reproduce the solution \cite{IKKMSS06}, see also \cite{CGI98}:
\begin{align}
 {\tilde V}_{x,\mu}
 = U_{x,\mu}  
 +  4{\bm n}_{x} U_{x,\mu} {\bm n}_{x+\mu}
  .
 \label{SU2-sol}
\end{align}

For $SU(3)$, for $\ell=1$, (\ref{nVU}) is 
\begin{align}
0  =     \alpha_1 
+ \sum_{j=1}^{2} \sum_{k=1}^{2} \frac{1}{4}\gamma_{jk} 
 {\rm tr}(({\bm n}^1_{x}*{\bm n}^j_{x}) U_{x,\mu} {\bm n}^k_{x+\mu}U_{x,\mu}^\dagger) 
 =  \alpha_1  
+  \frac{1}{4}\gamma_{12} 
 \frac{1}{\sqrt{3}}\frac{1}{2}
+  \frac{1}{4}\gamma_{21} 
 \frac{1}{\sqrt{3}}\frac{1}{2}
  =   \frac{3}{2} \alpha_1 
   ,
\end{align}
where we have used
(\ref{nSU3}), (\ref{SU3-parameter}) and (\ref{tr(nUnU)}).
Similarly, for $\ell=2$, (\ref{nVU}) is 
\begin{align}
0  =&    \alpha_2 
+ \sum_{j=1}^{2} \sum_{k=1}^{2} \frac{1}{4}\gamma_{jk} 
 {\rm tr}(({\bm n}^2_{x}*{\bm n}^j_{x}) U_{x,\mu} {\bm n}^k_{x+\mu}U_{x,\mu}^\dagger) 
=   \frac{3}{2} \alpha_2
 .
\end{align}
For $SU(3)$, thus, we have determined the parameters as
\begin{align}
 \alpha_1=\beta_1 = \alpha_2=\beta_2 = 0  , \quad 
 \gamma_{11} =  \gamma_{22} = 6 , \quad \gamma_{21} = \gamma_{12} = 0
 ,
\end{align}
and have obtained the solution of the defining equations:
\begin{align}
{\tilde V}_{x,\mu}  = U_{x,\mu} +  6{\bm n}^1_{x}U_{x,\mu}{\bm n}^1_{x}  +  6{\bm n}^2_{x}U_{x,\mu}{\bm n}^2_{x}
   .
\end{align}

Thus we have shown that the solution for $\tilde{V}_{x,\mu}$ is given for $N=2,3$ in the form:
\begin{align}
 \tilde{V}_{x,\mu}
= U_{x,\mu} + \sum_{j=1}^{N-1} 2N {\bm n}^j_{x} U_{x,\mu} {\bm n}^j_{x+\mu}  
  .
  \label{Vtilde}
\end{align}
It should be remarked that (\ref{Vtilde}) is also a solution for $SU(N)$ with $N >3$, since it is possible to check that (\ref{Vtilde}) fulfils the defining equations (a') and (b') for any $N$ of $SU(N)$.
Furthermore, it is possible to prove that it is a unique solution of the defining equations which is slightly modified for (b).  This result will be published elsewhere, because of limitations of space. 

The unitary link variable $V_{x,\mu}$ is obtained  from ${\tilde V}_{x,\mu}$ (\ref{Vtilde})  following  (\ref{polar}) and (\ref{speciality}).
Moreover, it is not difficult to show that the lattice condition (b')
$
 {\rm tr}({\bm n}_{x}^\ell \tilde{X}_{x,\mu}) 
=    {\rm tr}({\bm n}_{x}^\ell (\tilde{V}_{x,\mu}U_{x,\mu}^{-1})^{-1} ) = 0 
$
reproduces the continuum version 
$
2 {\rm tr}({\bm n}_{x}^\ell \mathscr{X}_\mu(x) )
={\bf n}^\ell(x) \cdot \mathbb{X}_\mu(x) =0
$ in the naive continuum limit. 

\section{Minimal case}

Now we discuss the minimal case  in which   $\mathscr A_\mu$ is decomposed into 
$\mathscr{V}_\mu(x)$ and $\mathscr{X}_\mu(x)$, i.e.,  
$
 \mathscr{A}_\mu(x) = \mathscr{V}_\mu(x) + \mathscr{X}_\mu(x) 
$, 
using a single color  field  
\begin{align}
 \bm h(x):=\bm n_r(x) =U^\dagger(x) H_r U(x)
  ,
\end{align}
where the final Cartan matrix is given by
$
 H_r  
 = \frac{1}{\sqrt{2N(N-1)}} {\rm diag}(1,\cdots,1,-N+1) 
 .
$

\subsection{Continuum: minimal case}

The respective components $\mathscr V_\mu(x)$ and $\mathscr X_\mu(x)$ are specified by two defining equations:

\noindent
(I)  $\bm{h}(x)$ is covariantly constant in the background $\mathscr{V}_\mu(x)$:
\begin{subequations}
\begin{align}
  0 = \mathscr{D}_\mu[\mathscr{V}] \bm{h}(x) 
:=\partial_\mu \bm{h}(x) -  ig [\mathscr{V}_\mu(x), \bm{h}(x)]
 ,
\label{defVL2}
\end{align}
(II)  $\mathscr{X}^\mu(x)$  has the vanishing $\tilde{H}$-commutative part, i.e., $\mathscr{X}^\mu(x)_{\tilde{H}}=0$:
\begin{align}
  \mathscr{X}^\mu(x)_{\tilde{H}} := \left( {\bf 1} -   2\frac{N-1}{N}  [\bm{h} , [\bm{h} ,  \cdot ]]
\right) \mathscr{X}_\mu(x)   = 0  
\label{defXL2}
 . 
\end{align}
\end{subequations}
By solving the defining equations, it has  been shown \cite{KSM08} that the new variables are written in terms of $\bm h$ and $\mathscr A_\mu$: 
\begin{subequations}
\begin{align}
\mathscr A_\mu(x)
 =& \mathscr V_\mu(x)
  +\mathscr X_\mu(x)
 =\mathscr C_\mu(x)
  +\mathscr B_\mu(x)
  +\mathscr X_\mu(x)
 ,
\\
  \mathscr{C}_\mu(x)
=& \mathscr{A}_\mu(x) - \frac{2(N-1)}{N}   [\bm{h}(x), [ \bm{h}(x), \mathscr{A}_\mu(x) ] ]
,
\\
 \mathscr{B}_\mu(x)
=& ig^{-1} \frac{2(N-1)}{N}[\bm{h}(x) , \partial_\mu  \bm{h}(x) ]
,
\\
 \mathscr{X}_\mu(x) =& -ig^{-1}  \frac{2(N-1)}{N}  [\bm{h}(x), \mathscr{D}_\mu[\mathscr{A}]\bm{h}(x) ]
 ,
\end{align}
 \label{NLCV-minimal}
\end{subequations}
once a single color field $\bm{h}(x)$ is given as a functional of $\mathscr A_\mu$.

\subsection{Lattice: minimal case}

In order to obtain the defining equations on a lattice and to solve them, we  repeat the same argument as that in the maximal case.
Consequently, the lattice versions  of the defining equations in the minimal choice for $SU(N)$ are  given by
\footnote{
Note that (\ref{cond2m-min}) is sufficient to determine the ansatz ${\tilde V}_{x,\mu}$ of the form (\ref{ansatzmini}), although we could impose a stronger condition on the lattice which is equivalent to (\ref{defXL2}) in the continuum form. 
}
\begin{subequations}
\begin{align}
 & \text{(a')} \quad 
 \bm{h}_{x} V_{x,\mu}  = V_{x,\mu} \bm{h}_{x+\mu}   
  ,
 \label{Lcc-min}
\\
 & \text{(b')} \quad
 {\rm tr}(\bm{h}_{x} U_{x,\mu} V_{x,\mu}^\dagger) 
    =  0 = {\rm tr}(\bm{h}_{x} V_{x,\mu} U_{x,\mu}^\dagger) 
 .
  \label{cond2m-min}
\end{align}
\end{subequations}


In the minimal case, we adopt the ansatz for $V_{x,\mu}$:
\begin{align}
 {\tilde V}_{x,\mu}
 = U_{x,\mu} +  \alpha  {\bm h}_{x} U_{x,\mu} +  \beta U_{x,\mu} {\bm h}_{x+\mu} 
 +  \gamma {\bm h}_{x} U_{x,\mu} {\bm h}_{x+\mu}
  .
 \label{ansatzmini}
\end{align}
Then we calculate
\begin{align}
 {\bm h}_{x} {\tilde V}_{x,\mu}
 =&   \frac{\alpha }{2N}  U_{x,\mu} 
+ \left( 1 +  \frac{\kappa}{2}  \alpha  \right) {\bm h}_{x} U_{x,\mu}
 +   
 \frac{\gamma}{2N} U_{x,\mu} {\bm h}_{x+\mu}
+ \left( \beta +  \frac{\kappa}{2}  \gamma  \right)  {\bm h}_{x} U_{x,\mu} {\bm h}_{x+\mu} 
 ,
\nonumber\\
 {\tilde V}_{x,\mu} {\bm h}_{x+\mu} 
=& \frac{\beta}{2N}   U_{x,\mu}  
+ 
 \frac{\gamma}{2N}  {\bm h}_{x} U_{x,\mu} 
+  \left( 1 +  \frac{\kappa}{2}  \beta  \right)  U_{x,\mu} {\bm h}_{x+\mu} 
+  \left( \alpha +  \frac{\kappa}{2}  \gamma  \right)  {\bm h}_{x} U_{x,\mu} {\bm h}_{x+\mu} 
  ,
 \label{ansatz2a}
\end{align}
where we have used
$
  {\bm h}_{x} {\bm h}_{x}
=  \frac{1}{2N}  {\bf 1} + (\kappa/2) {\bm h}_{x}
$
with $\kappa :=\frac{2(2-N)}{\sqrt{2N(N-1)}}$.
By comparing the right-hand sides of two equations,  
it turns out that (a') is satisfied when 
\begin{equation}
\alpha=\beta,
\quad
1+(\kappa/2) \alpha=\gamma/(2N).
\label{eq:parameter-eq1}
\end{equation}

Moreover, we calculate 
\begin{align}
{\bm h_x}{\tilde V}_{x,\mu}U_{x,\mu}^\dagger
 &=\frac\alpha{2N}\bm1
   +\left(1+\frac\kappa2\alpha\right){\bm h}_x 
   +\frac\gamma{2N}U_{x,\mu}{\bm h}_{x+\mu}U_{x,\mu}^\dagger
   \nonumber\\
 &\qquad
   +\left(\beta+\frac\kappa2\gamma\right){\bm h}_x U_{x,\mu}{\bm h}_{x+\mu}U_{x,\mu}^\dagger
    .
\end{align}
Therefore,  (b')
  yields 
\begin{equation}
 {\rm Tr}({\bm h}_x {\tilde V}_{x,\mu}U_{x,\mu}^\dagger)
= \alpha+ (\kappa/4) \gamma=0
 ,
\label{eq:parameter-eq2}
\end{equation}
where we have used $\alpha=\beta$, 
${\rm Tr}({\bf 1})=N$,
${\rm Tr}({\bm h}_x)=0$,
and ${\rm Tr}({\bm h_x}U_{x,\mu}{\bm h}_{x+\mu}U_{x,\mu}^\dagger)=1/2$.
Solving (\ref{eq:parameter-eq1})
 and (\ref{eq:parameter-eq2})
 simultaneously, we obtain
\begin{equation}
\alpha=\beta=\frac{N-2}{N^2-2N+2}\sqrt{2N(N-1)},
\quad
\gamma=\frac{4N(N-1)}{N^2-2N+2}
 .
\label{eq:parameters}
\end{equation}
In fact, the ansatz (\ref{ansatzmini}) for ${\tilde V}_{x,\mu}$ correctly reproduces the continuum counterpart $\mathscr{V}_\mu(x)$ in the naive continuum limit.
\begin{align}
 {\tilde V}_{x,\mu}
 =& \left(1+\frac\gamma{2N}\right)\bm1
   -i\epsilon g[{\mathscr A_{x,\mu}}+\alpha\{{\bm h_x},{\mathscr A_{x,\mu}}\}+\gamma{\bm h_x}{\mathscr A_{x,\mu}}{\bm h_x}]
   +\epsilon[\alpha\partial_\mu{\bm h_x}+\gamma{\bm h_x}\partial_\mu{\bm h_x}]
+O(\epsilon^2)
\nonumber\\
 =&  \left(1+\frac\gamma{2N}\right) [\bm1-i \epsilon g \mathscr{V}_\mu(x) +O(\epsilon^2) ] 
  .
\end{align}

For $SU(2)$, $\alpha=\beta=0$ and the linear terms in ${\bm h}$ vanish and (\ref{ansatzmini}) reduces to the same form (\ref{SU2-sol}) as the maximal case $\gamma=4$.  This is reasonable, since there is no distinction between maximal and minimal for $SU(2)$. 

For $SU(3)$, the parameters are decided as 
\begin{equation}
\alpha=(2/5) \sqrt3,
\quad
\gamma=24/5
 ,
\end{equation}
which leads to the link variable (up to the normalization):
\begin{equation}
{\tilde V}_{x,\mu}=U_{x,\mu}+\frac25\sqrt3({\bm h_x}U_{x,\mu}+U_{x,\mu}\bm h_{x+\mu})+\frac{24}5{\bm h_x}U_{x,\mu}\bm h_{x+\mu}
 .
\end{equation}
In particular, the large $N$ limit $N \rightarrow \infty$ is taken in the minimal case:
\begin{equation}
{\tilde V}_{x,\mu}=U_{x,\mu}+ \sqrt2({\bm h_x}U_{x,\mu}+U_{x,\mu}\bm h_{x+\mu})+4{\bm h_x}U_{x,\mu}\bm h_{x+\mu}
 .
 \label{tildeV2}
\end{equation}
The unitary link variable $V_{x,\mu}$ is obtained from ${\tilde V}_{x,\mu}$ (\ref{tildeV2}) following  (\ref{polar}) and (\ref{speciality}).



\section{Construction of the color field}

Now we discuss how to obtain the desired color field $\bm{n}_x$ from the original lattice Yang-Mills theory with a gauge group $G=SU(N)$ written in terms of link variables $U_{x,\mu}$. 
Such a prescription has already been given in the continuum reformulation of Yang-Mills theory \cite{KMS06,KSM08}. 
Here we present the lattice version.


We can introduce  a single color field  $\bm{n}_x$  according to $\bm{n}_x=U_x^\dagger T U_x$ ($U_x \in G$) for a certain diagonal matrix $T$ \cite{KSM08} in addition to link variables $U_{x,\mu}$. Then we have an extended gauge theory, 
called the master Yang-Mills theory in which fields  transform under the enlarged gauge transformation:
the color field transforms as 
\begin{align}
  {\bm n}_{x} \rightarrow \Theta_{x} {\bm n}_{x} \Theta_{x}^\dagger = {\bm n}_{x}' ,  \quad 
 \Theta_{x} \in G/\tilde{H}
\label{n-transf2}
 ,
\end{align}
when the link variable $U_{x,\mu}$ obeys the well-known lattice gauge transformation:
\begin{equation}
  U_{x,\mu}  \rightarrow \Omega_{x} U_{x,\mu} \Omega_{x+\mu}^\dagger = U_{x,\mu}^\prime
  , \quad \Omega_{x} \in G 
  \label{U-transf2}
 .
\end{equation}
Therefore, the master Yang-Mills theory has the enlarged gauge symmetry $\tilde{G}:=G_{\Omega} \times (G/\tilde{H})_{\Theta}$.
Here $G/\tilde{H}$ denotes the target space of the color field: $\bm{n}_x \in G/\tilde{H}$ and the subgroup $\tilde{H}$ depends on the chosen matrix $T$. 
The maximal case  corresponds to $\tilde{H}=U(1)^{N-1}$, while the minimal case  to $\tilde{H}=U(N-1)$.
For $SU(2)$, in particular, a single color field is unique, and $\tilde H$ agrees with the maximal torus subgroup: $\tilde H=H=U(1)$ and $\bm n_x \in SU(2)/U(1)$.

\begin{figure}[ptb]
\begin{center}
\includegraphics[height=2.9in, angle=-90]{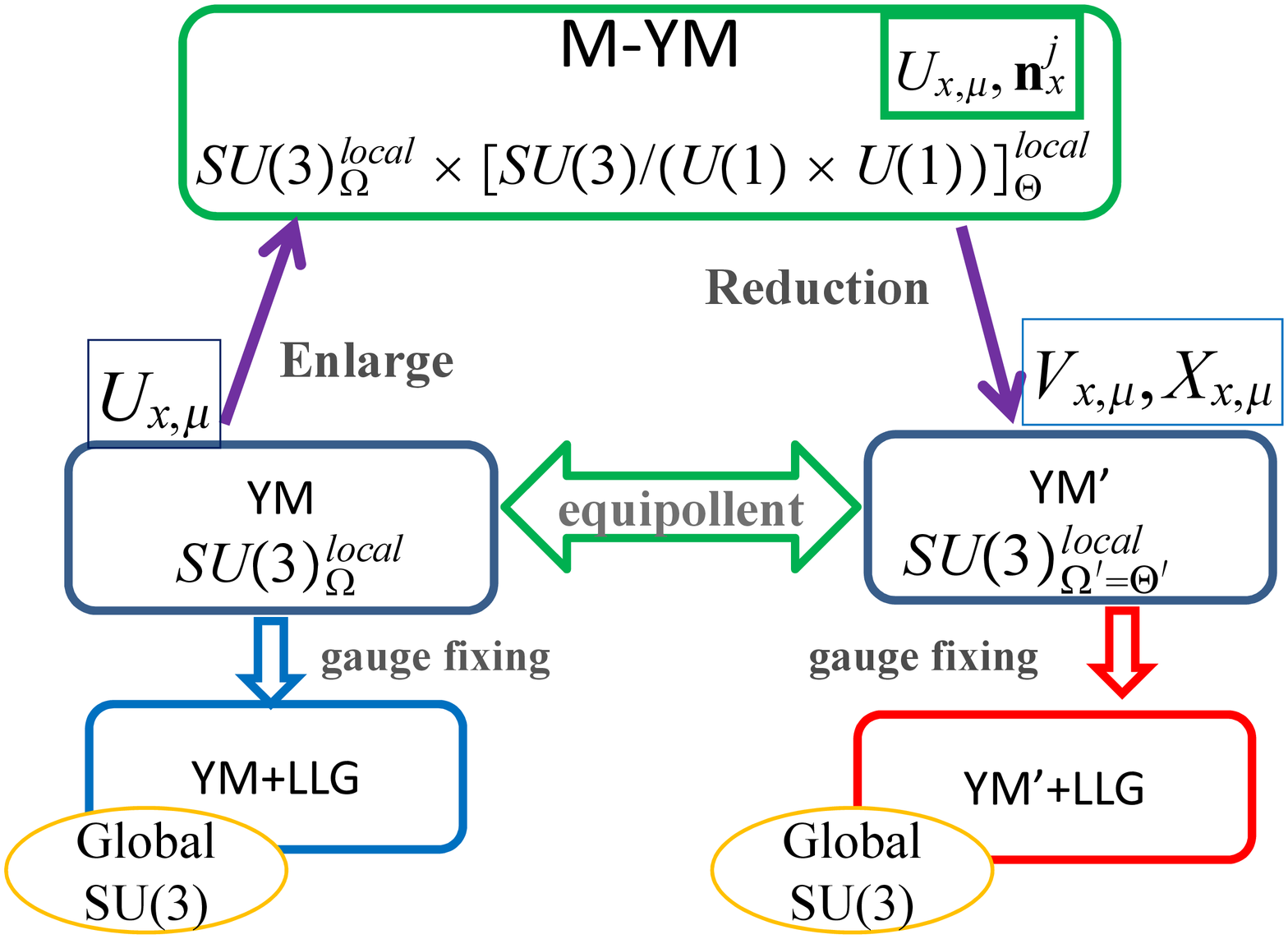}
\includegraphics[height=2.9in, angle=-90]{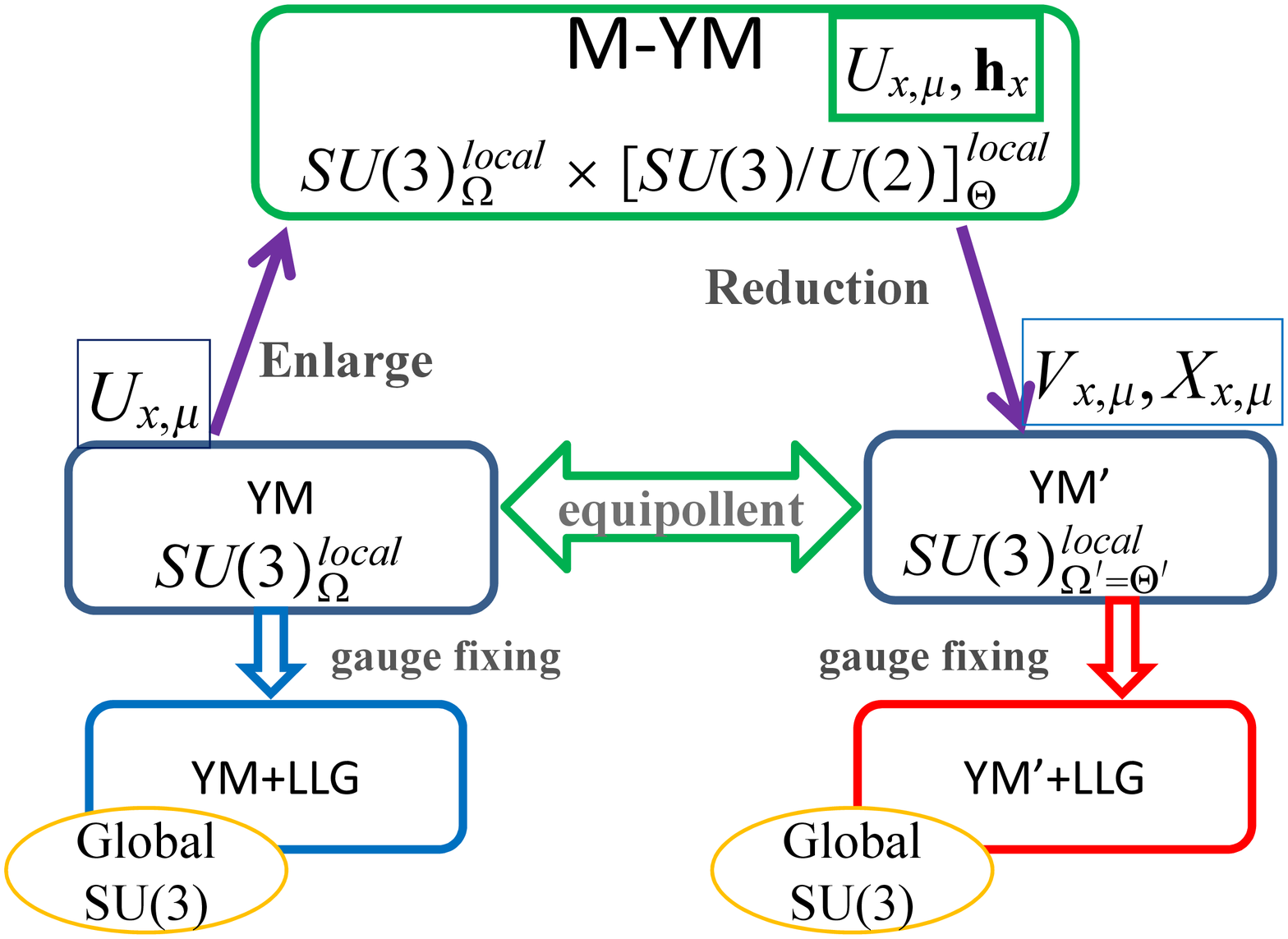}
\end{center}
\caption{The original Yang-Mills (YM) theory with a gauge group $G_{\Omega}$ is first extended to the master Yang-Mills theory (M-YM) with the enlarged gauge symmetry $\tilde{G}:=G_{\Omega} \times (G/\tilde{H})_{\Theta}$ and subsequently reduced by imposing reduction conditions to the equipollent Yang-Mills (YM') theory with $G^\prime_{\Omega^\prime}$. The overall gauge fixing condition e.g., Lattice Landau gauge (LLG) can be imposed without breaking color symmetry. 
(Left panel) SU(3) maximal case. (Right panel) SU(3) minimal case. 
}%
\label{fig:MYM}%
\end{figure}

However, the master Yang-Mills theory has more degrees of freedom than the original Yang-Mills theory by construction. 
We impose a constraint called the reduction condition (mimicking the continuum version \cite{KMS06,KSM08}) to reduce 
$\tilde{G}:=G_{\Omega} \times (G/\tilde{H})_{\Theta}$ in the master Yang-Mills theory to the subgroup: $G^\prime=SU(N)_{\Omega^\prime}$:
\begin{equation}
 G=SU(N)^{\rm local}_{\Omega} 
 \nearrow 
\tilde{G}:=SU(N)^{\rm local}_{\Omega} \times [SU(N)/\tilde{H}]^{\rm local}_{\Theta} 
 \searrow 
G^\prime :=SU(N)^{\rm local}_{\Omega^\prime}
 ,
\end{equation}
so that the resulting theory has the same local gauge symmetry $G'$ (equipollent gauge symmetry) as the original Yang-Mills theory. 
Thus we have a new description of Yang-Mills theory for a choice of $T$. 
See Fig.~\ref{fig:MYM}.

This procedure should be compared with the conventional MAG. The conventional MAG  is the partial gauge fixing $G^{\rm local} \rightarrow H^{\rm local}$ which breaks the local gauge symmetry $G^{\rm local}$ to  a maximal torus group $H^{\rm local}=U(1)^{N-1}$  and leaves both local $H^{\rm local}=U(1)^{N-1}$ and  global $H^{\rm global}=U(1)^{N-1}$ unbroken, but breaks the global $SU(N)^{\rm global}$, i.e., color symmetry.
The reduction condition leaves both local $G'^{\rm local}=SU(N)'$ and global $SU(N)'^{\rm global}$ unbroken (color rotation invariant). 
In order to fix the local gauge symmetry $G'^{\rm local}$, we can impose an overall gauge fixing condition (e.g., Landau gauge) with the color symmetry being preserved.
This is an advantage of the new procedure.

In numerical simulations, 
 the configurations of $SU(N)$ link variables 
$\{ U_{x,\mu} \}$ are generated using the standard procedure for the  Wilson action \cite{CM82}.
For $SU(3)$ 
the distinction between the maximal and minimal cases is attributed to the method of constructing the color field by imposing appropriate reduction conditions given below.


In the maximal case, the variables $\bm{n}^j_{x}$ are constructed as follows. 
Consider  two functionals $F_{\rm rc:max}$ and $F_{LLG}$ written in terms of   $U_{x,\mu}$ and $\bm{n}^j_{x}$:  
\begin{align}
 F_{\rm rc:max}[U,\bm{n};\Omega,\Theta]
 :=& \epsilon^D \sum_{x,\mu,j} \frac12 {\rm tr}\left[ (D_\mu^{(\epsilon)}[{}^{\Omega}U]{}^{\Theta}\bm{n}^j_x)^\dagger (D_\mu^{(\epsilon)}[{}^{\Omega}U]{}^{\Theta}\bm{n}^j_x)\right] 
\nonumber\\
=& - \sum_{x,\mu,j} \epsilon^{D-2}  {\rm tr} 
  \left[  {}^{\Theta}\bm{n}_{x}^{j} {}^{\Omega}U_{x,\mu} {}^{\Theta}\bm{n}_{x+\mu}^{j}  {}^{\Omega}U_{x,\mu}^\dagger -\frac{N-1}{2}    \right]
   ,
\label{Lattice-nMAGf}
\\
F_{LLG}[U;\Omega]
:=& - \epsilon^{D}  \sum_{x,\mu}\mathrm{tr}( {}^{\Omega} U_{x,\mu}) 
 .
\end{align}
Then we  minimize simultaneously the two functionals $F_{\rm rc:max}$ and $F_{LLG}$ with respect to two gauge transformations: 
${}^{\Omega}{}U_{x,\mu}:=\Omega_{x}U_{x,\mu}\Omega_{x+\mu}^{\dagger}$ for the
link variable $U_{x,\mu}$ and 
${}^{\Theta}\bm{n}^j_{x}:=\Theta_{x} \bm{n}_{x}^{j(0)}\Theta_{x}^{\dagger}$ for a given initial site variable
$\bm{n}^j_{x}{}^{(0)}$ (we can choose the initial value $\bm{n}^j_{x}{}^{(0)}=H_{j}$)
where  gauge group elements $\Omega_{x}$ and $\Theta_{x}$
are {\it independent} $SU(N)$ matrices on a site $x$. 
Then we can determine  the configurations  ${}^{\Theta^{*}}\bm{n}^j_{x}$ and ${}^{\Omega^{*}}U_{x,\mu}$ realizing simultaneously the minimum of two functionals, just as in the $SU(2)$ case \cite{IKKMSS06}.

It is easy to show that the continuum limit of the lattice functional (\ref{Lattice-nMAGf}) reduces to the continuum one in \cite{KSM08}:
Expanding the link variable in power series of $\epsilon$, we obtain
\begin{align}
 F_{\rm rc:max}[U, \bm{n}] 
  =&  \epsilon^D  \sum_{x,\mu,j} \frac12 {\rm tr} \{ (D_\mu[\mathscr{A}] \bm{n}_{x}^{j} )^2 \} 
  + O(\epsilon^2)  
 ,
\end{align}
where  we have used
$
  \bm{n}_{x+\mu}^{j} = \bm{n}_{x}^{j} + \epsilon \partial_\mu \bm{n}_{x}^j + O(\epsilon^2) ,
$
and 
$
 {\bm n}^j_{x} {\bm n}^j_{x} 
 = \frac{N-1}{2N} {\bf 1}
$.
Indeed, this result reproduces the continuum functional in the maximal case\cite{KSM08}:
$
R= \int d^Dx \frac12 (D_\mu[\mathscr{A}]\bm{n}^j(x))^2 
  .
$
There is the other choice for the $F_{\rm rc:max}[U, \bm{n}]$, see  \cite{KSM08} .


The minimal case is achieved by minimizing two functionals simultaneously:
\begin{align}
 F_{\rm rc:min}[U,\bm{h};\Omega,\Theta]
 :=& \epsilon^D \sum_{x,\mu,j} \frac12 {\rm tr}\left[ (D_\mu^{(\epsilon)}[{}^{\Omega}U]{}^{\Theta}\bm{h}_x)^\dagger (D_\mu^{(\epsilon)}[{}^{\Omega}U]{}^{\Theta}\bm{h}_x)\right]  
 \nonumber\\
  =& - \sum_{x,\mu} \epsilon^{D-2} {\rm tr} 
 \left[  {}^{\Theta}\bm{h}_{x}  {}^{\Omega}U_{x,\mu} {}^{\Theta}\bm{h}_{x+\mu}  {}^{\Omega}U_{x,\mu}^\dagger  -\frac{1}{2}  \right] 
 ,
\\
F_{LLG}[U;\Omega]
:=& - \epsilon^{D-4} \sum_{x,\mu}\mathrm{tr}({}^{\Omega} U_{x,\mu}) .
\end{align}
The naive continuum limit of this functional is calculated as in the maximal case:
\begin{align}
   F_{\rm rc:min}[U, \bm{h}] 
  =&  \epsilon^{D}  \sum_{x,\mu}   \left[  \frac12 \epsilon^2 {\rm tr} \{ (D_\mu[\mathscr{A}] \bm{h}_{x} )^2 \}
\right] 
  + O(\epsilon^2)  
   ,
\end{align}
where we have used  
$
 H_r H_r
 = \frac{1}{2N(N-1)} {\rm diag}(1,\cdots,1,(N-1)^2) 
$
and 
$
  \bm{h}_{x+\mu}  = \bm{h}_{x}  + \epsilon \partial_\mu \bm{h}_{x}  + O(\epsilon^2) 
$.
Indeed, this result reproduces the continuum functional in the minimal case \cite{KSM08}:
$
 F = \int d^Dx \frac12 (D_\mu[\mathscr{A}]\bm{h}(x))^2 
$
 apart from a constant term.
More practical procedures necessary in numerical simulations will be given in subsequent papers.

\section{Conclusion and discussion}

In this Letter we have given new descriptions for lattice Yang-Mills theory in terms of new variables. 
We expect that they are quite useful to give a better understanding of dual superconductivity and to clarify the true mechanism for quark confinement in $SU(N)$ Yang-Mills theory. 
In fact, some applications of this framework were already presented in the continuum formulation \cite{Kondo08,KS08,Kondo08b}.  
The results of relevant numerical simulations on a lattice will be reported soon, although  preliminary results were given in  \cite{SKKMSI07b} for $SU(3)$ maximal choice.    

We have presented a set of defining equations for specifying the decomposition of lattice variables to obtain new lattice variables. However, it is not the unique way for the decomposition. 
The first defining equation for $V_{x,\mu}$ has the intrinsic meaning even on a lattice without referring to the continuum limit, since it implies that the color field $\bm n_x$ at a site $x$ can be moved to the next site $x+\mu$ without being affected by the background field expressed as the link variable $V_{x,\mu}$. 
On the other hand, the second defining equation (the extra condition) for $X_{x,\mu}$  will not be the best form, although it is enough to reproduce the continuum form expressed in the Lie algebra, 
${\rm tr}(\bm{n}(x) \mathscr{X}_\mu(x))=0$. 
This issue will be discussed in a separate paper.

\appendix
\section{Spectral resolution}\label{appendix:spectral-resolution}

A square matrix $M$ is called a normal matrix, if $MM^\dagger=M^\dagger M$. 
The Toeplitz theorem tells us that a complex square matrix $M$ can be diagonalized by a unitary matrix if and only if $M$ is a normal matrix.   In particular,  Hermitian matrices and unitary matrices are examples of the normal matrix. 
For a normal matrix $M$, let $\lambda_1, \cdots, \lambda_n$ be the non-coinciding eigenvalues. Then $M$ has the unique spectral resolution with projection operators $\mathscr{P}_1,  \cdots , \mathscr{P}_n$:
\begin{align}
 M =& \lambda_1 \mathscr{P}_1 + \cdots + \lambda_n \mathscr{P}_n ,
   \quad
 \mathscr{P}_1 + \cdots +  \mathscr{P}_n  = I , 
\nonumber\\
   & \mathscr{P}_j  \mathscr{P}_j  = \mathscr{P}_j , 
   \quad
    \mathscr{P}_j^\dagger = \mathscr{P}_j ,
   \quad
    \mathscr{P}_j  \mathscr{P}_k  = 0 \ (j \ne k) 
    .
    \label{spec-rep}
\end{align}
Conversely, if there exist such projection operators satisfying the above properties, the matrix $M$ becomes the normal matrix. 
The projection operators are constructed as follows.
Let $\bm{x}_j$ be the eigenvector (represented by a column vector) corresponding to the eigenvalue $\lambda_j$, i.e., $M \bm{x}_j=\lambda_j \bm{x}_j$,   
and the eigenvectors satisfy $\bm{x}_j^\dagger \bm{x}_k=\delta_{jk}$. 
Define a matrix $U$ by 
$U:=(\bm{x}_1 \cdots \bm{x}_n)$.  Then $U$ is a unitary matrix and 
$MU=U{\rm diag}(\lambda_1, \cdots, \lambda_n)$.
Therefore, $M$ can be diagonalized by a unitary matrix $U=(\bm{x}_1 \cdots \bm{x}_n)$ as $U^\dagger M U={\rm diag}(\lambda_1, \cdots, \lambda_n)$. This implies that 
$M=U {\rm diag}(\lambda_1, \cdots, \lambda_n) U^\dagger
= \sum_{j=1}^{n} \lambda_j \bm{x}_j \bm{x}_j^\dagger
= \sum_{j=1}^{n} \lambda_j \mathscr{P}_j$ with matrices $\mathscr{P}_j=\bm{x}_j \bm{x}_j^\dagger$.
We can check that these $\mathscr{P}_j$ satisfy the above properties of the projection operators. 

For a matrix $A$ to be commutable with $M$, it is necessary and sufficient that every $\mathscr{P}_j$ ($j=1, \cdots, n$) commutes with $A$:
\begin{equation}
 [M, A] =0 \Longleftrightarrow [\mathscr{P}_j ,A]=0 \ ( j=1, \cdots, n )
  .
\end{equation}
For a positive-definite Hermitian matrix $M$, there is a unique positive-definite matrix $L$ such that $M=L^m$. 
See \cite{Satake73} for the proofs of the above theorems. 

We apply this theorem to the positive definite Hermitian matrix $M=P^2=\tilde V \tilde V^\dagger$ whose eigenvalues are real positive $\lambda_1, \cdots, \lambda_n>0$.  $P^2$ has the spectral resolution (\ref{spec-rep}). 
Then $P$ is also a positive definite Hermitian matrix.  
The Hermitian matrix is a normal matrix and has the spectral resolution
\begin{align}
 P = \sqrt{\lambda_1} \mathscr{P}_1 + \cdots + \sqrt{\lambda_n} \mathscr{P}_n  
    .
\end{align} 
Thus we conclude that if $[P^2, A] =0$, then $[P, A] = 0$. 
\begin{equation}
 [P^2, A] =0 \Longleftrightarrow [\mathscr{P}_j ,A]=0 \ ( j=1, \cdots, n ) \Longleftrightarrow [P, A] = 0 
  .
\end{equation}

Note that $P^{-1}$ exists. It is easy to show that
\begin{equation}
 [P, A] =0 \Longleftrightarrow [P^{-1} ,A]=0 
  .
\end{equation}
Thus we have shown the equivalence:
\begin{equation}
 [P^2, A] =0 \Longleftrightarrow  [P, A] =0 \Longleftrightarrow [P^{-1} ,A]=0 
  .
\end{equation}
If we choose the color field $\bm{n}$ as a matrix $A$, then $[P^2,\bm{n}]=0$ from the first defining equation, 
\begin{equation}
 [P^2, \bm{n}] =0 \Longleftrightarrow  [P,\bm{n}] =0 \Longleftrightarrow [P^{-1} ,\bm{n}]=0 
  .
\end{equation}

\section*{Acknowledgments}
This work is financially supported by Grant-in-Aid for Scientific Research (C) 18540251  from Japan Society for the Promotion of Science (JSPS) and the Large Scale Simulation Program No.07-15 (FY2007) of High Energy Accelerator Research Organization(KEK).


\end{document}